\documentclass[12pt]{article}

\usepackage{cite}
\usepackage{graphicx}
\usepackage{epsfig}
\usepackage{amsfonts}
\usepackage{amssymb}
\usepackage{amsmath,amssymb}
\usepackage{enumitem} 

\usepackage{epsf,epsfig}

\date{\today}

\newcommand{\ee}{\end{equation}}
\newcommand{\eea}{\end{eqnarray}}
\newcommand{\be}{\begin{equation}}
\newcommand{\bea}{\begin{eqnarray}}

\begin{document}
\begin{center}

{\Large \bf Black hole scalarisation 
\\ from the breakdown of scale-invariance}
\vspace{0.8cm}
\\
Carlos A. R. Herdeiro$^{\dagger}$ 
and  
 Eugen Radu$^{\ddagger}$ 
\vspace{0.3cm}
\\
$^{\dagger}${\small Centro de Astrof\'\i sica e Gravita\c c\~ao - CENTRA,} \\{\small Departamento de F\'\i sica,
Instituto Superior T\'ecnico - IST, Universidade de Lisboa - UL,} \\ {\small Avenida
Rovisco Pais 1, 1049-001, Portugal}
\vspace{0.3cm}
\\
$^{\ddagger }${\small  Center for Research and Development in Mathematics and Applications (CIDMA),
\\
Department of Mathematics,  Campus de Santiago, 3810-183 Aveiro, Portugal} 
\end{center}
\begin{abstract}
Electro-vacuum black holes are scale-invariant; their energy-momentum tensor is traceless. Quantum corrections of various sorts, however, can often produce a trace anomaly and a breakdown of scale-invariance. The (quantum-corrected) black hole solutions of the corresponding gravitational effective field theory (EFT) have a non-vanishing Ricci scalar. Then, the presence of a scalar field with the standard non-minimal coupling $\xi \phi^2 R$ naturally triggers a spontaneous scalarisation of the corresponding black holes.  This scalarisation phenomenon occurs for an  
(infinite) discrete set of $\xi$. We illustrate the occurrence of this phenomenon for two examples of static, spherically symmetric, asymptotically flat
 black hole solution of EFTs. In one example the trace anomaly comes from the matter sector -- a novel, closed form, generalisation of the Reissner-Nordstr\"om solution 
with an $F^4$ correction -- whereas in the other example it comes from the geometry sector --  a noncommutative geometry generalization of the Schwarzschild
black hole. For comparison, we also consider the scalarisation of a black hole surrounded by (non-conformally invariant) classical matter (Einstein-Maxwell-dilaton black holes). 
We find that the scalarised solutions are, generically, entropically favoured.
\end{abstract}

\tableofcontents

\section{Introduction and motivation}
The Schwarzschild black hole (BH) is scale-invariant. That is, even though vacuum General Relativity introduces a scale, the Planck length, $\ell_P$, via Newton's constant, a classical Schwarzschild BH with a Schwarzschild radius $R_S$ of the order of the Planck length $R_S\sim \ell_P$ and another with $R_S\sim 10^9 M_\odot$ (like to one at the centre of M87~\cite{Macchetto:1997gi})  are identical, up to a scale transformation. This scale invariance remains for all electro-vacuum BH solutions of General Relativity (the Kerr-Newman  family~\cite{Kerr:1963ud,Newman:1965my,Chrusciel:2012jk}). That is, fixing all other dimensionless parameters, Kerr-Newman BHs with different masses are mapped to one another by a scale transformation. Mathematically, scale-invariance is manifest in the tracelessness of the energy-momentum tensor.\footnote{It is pedagogical to contrast this scale-invariance with the scale-dependence of other solutions. Consider, for instance boson stars~\cite{Schunck:2003kk}, which are solutions of the Einstein-(complex-)Klein-Gordon model, with a mass term for the scalar field. Boson stars are not scale invariant and have qualitatively different features depending on their total mass. For instance, fixing all other parameters, there can be solutions which are very compact and have a light ring, or less compact solutions without a light ring~\cite{Cunha:2015yba}.}

This classical symmetry (scale invariance) may become anomalous at the quantum level. For instance, the electromagnetic sector is known to be conformal only classically (and in four spacetime dimensions), by virtue of the running of the coupling induced by the quantum vacuum  polarisation~\cite{Peskin:1995ev}. At the level of some effective field theory that takes into account the leading quantum effects ($e.g.$ Euler-Heisenberg non-linear electrodynamics~\cite{Heisenberg:1935qt}), this anomaly is materialised in the appearence of an energy-momentum tensor trace, which generically implies, via the semi-classical Einstein equations, a non-vanishing Ricci scalar.  Likewise, most approaches to quantum gravity will introduce a new scale as a cut off for the validity of the classical geometry, say, the string length in string theory or a non-commutativity parameter in non-commutative geometry (NCG). Generically, this scale will lead to the breakdown of scale invariance in the quantum corrected BH solutions and a non-vanishing Ricci scalar.  

Apart from a possible quantum origin, classical BHs with a non-vanishing Ricci scalar are also possible beyond electro-vacuum. This occurs when non-conformal matter is present in the action. An example will be presented below. The purpose of this paper is to consider a particular physical effect which can occur for BHs with a non-vanishing Ricci scalar:  the phenomenon of \textit{spontaneous scalarisation} of BHs due to a simple and well motivated non-minimal coupling between a scalar field and the Ricci scalar curvature. 

Spontaneous scalarisation of neutron stars has been discussed for over two decades, since the original proposal~\cite{Damour:1993hw}. It occurs in the context of scalar-tensor theories, wherein a scalar field can be sourced by the trace of the energy-momentum tensor. For neutron star geometries, this trace is non-vanishing and, in some regions of the parameter space, it  becomes energetically favoured for the neutron star to develop a scalar ``halo" around it, $i.e.$, to scalarise. Electro-vacuum BHs, on the other hand, have a vanishing Ricci scalar and cannot, therefore, source a scalar field and get scalarised in this context~\cite{Hawking:1972qk,Sotiriou:2011dz}. Still, a similar phenomenon was suggested to also occur for BHs if matter were present in the vicinity of the BH~\cite{Cardoso:2013fwa,Cardoso:2013opa}.  Then, Schwarzschild/Kerr BHs could be unstable against such spontaneous scalarisation.  A concrete realisation of this idea was presented in~\cite{Kleihaus:2015iea} wherein a Kerr BH with synchronised hair~\cite{Herdeiro:2014goa} was shown to have scalarised counterparts. In these examples, spontaneous scalarisation relies on the existence of a non-minimal coupling between the scalar field and the Ricci scalar and may only occur for backgrounds with non-vanishing Ricci scalar, which requires the presence of matter. Moving from the Jordan to the Einstein frame, moreover, the non-minimal coupling to the curvature disappears, and a non-minimal coupling to matter emerges. These observations justify the perspective that this phenomenon is a \textit{matter-induced spontaneous scalarisation}.  

On the other hand, a new guise of the spontaneous scalarisation phenomenon, dubbed \textit{geometric spontaneous scalarisation}, has recently been under scrutiny. In \cite{Doneva:2017bvd,Silva:2017uqg,Antoniou:2017acq} it was pointed out that in gravitational models where a real scalar field minimally couples to the curvature squared Gauss-Bonnet combination, under certain choices of the coupling function, both the standard (bald) vacuum BH solutions of general relativity and new ``hairy" BH solutions with a scalar field profile are possible, circumventing  no-scalar hair theorems~\cite{Herdeiro:2015waa} (see also~\cite{Kanti:1995vq} for earlier solutions). It was moreover suggested that the hairy BHs could form via spontaneous scalarisation, since the bald BH solutions were shown to be perturbatively unstable~\cite{Doneva:2017bvd,Silva:2017uqg}. Confirming this possibility, however, requires performing dynamical evolutions of the instability in the fully non-linear theory, which has not been achieved yet. But the suggestion that spontaneous scalarisation occurs dynamically for BHs could be confirmed in a cousin model~\cite{Herdeiro:2018wub}. In this class of models there are no non-minimal couplings between the scalar field and the curvature; there is a non-minimal coupling between the scalar and the electromagnetic field. Thus it falls in the class of matter-induced spontaneous scalarisation, where matter here is the electromagnetic energy.  But the chosen source term for the scalar field (the Maxwell invariant) does not require a non-vanishing Ricci scalar for a BH to scalarise; it requires electromagnetic charge.  
Scalarisation of this sort can occur both for a charged sphere in flat spacetime and for electrically charged BHs; thus gravity is optional. In this model, fully non-linear numerical simulations could be performed, showing the unstable bald Reissner-Nordstr\"om (RN) BHs grow scalar hair, and the growth saturates to match a hairy (or scalarised) solution ~\cite{Herdeiro:2018wub}. 
Subsequent related work both on geometric and matter-induced spontaneous scalarisation of BHs can be found in~\cite{Antoniou:2017hxj,Blazquez-Salcedo:2018jnn,Myung:2018vug,Doneva:2018rou,Brihaye:2018bgc,Boskovic:2018lkj,Minamitsuji:2018xde,Myung:2018jvi,Silva:2018qhn}.

A simple, often used in the context of quantum field theory in curved spacetime~\cite{Birrell:1982ix}, non-minimal coupling between curvature and a scalar field is of the form $\xi \phi^2 R$, where $R$ is the Ricci scalar. There are several motivations for this coupling that we shall review below. Since, as argued below, both quantum corrected BHs emerging within some effective field theory and classical solutions beyond electro-vacuum can have a non-vanishing Ricci scalar, here we study the possibility that spontaneous scalarisation exists due to this coupling. Within the quantum considerations we illustrate this possibility with two concrete examples. We shall see that indeed the non-scale invariant BHs are unstable against scalarisation and, moreover, that there are scalarised BH solutions. The latter are entropically preferred over the scalar-free ones in the model where entropy in unambiguous. These observations support the suggestion that spontaneous scalarisation occurs dynamically, even for BHs that would be classically scale invariant, once quantum corrections are taken into account. Considering also an example of a BH with classically non-scale invariant matter, provides evidence that scalarisation occurs universally for non-scale invariant BHs, regardless of the classical or quantum origin of the scale-invariance ``anomaly". 

This paper is organised as follows. In Section~\ref{sec2} we review the basics of the spontaneous scalarisation phenomenon for BHs, discuss the different types thereof and in particular the one that shall be addressed herein, also presenting the specific BH examples that shall be studied in detail.  In Section~\ref{sec3} we present the general formalism for obtaining both test field scalar clouds around BHs and the fully non-linear scalarised solutions. In Section~\ref{sec4}-\ref{sec6} we present the three illustrative examples of BH scalarisation we have already mentioned. Some conclusions are drawn in Section~\ref{sec7}.

\section{Scalarisation in a nutshell}
\label{sec2}
BH scalarisation occurs in models described by the Einstein-Hilbert action plus a scalar field action (and plus other possible matter terms). The scalar field action takes the generic form
\begin{eqnarray}
\label{actionS}
\mathcal{S}_\phi=- \int d^4 x \sqrt{-g} 
\left[
 \frac{1}{2}(\nabla \phi)^2 
+f(\phi) {\cal I}(\psi;g)
\right] \ ,
\end{eqnarray}
where $f(\phi) $ is the {\it coupling function} and  ${\cal I}$ is 
a source term which generically depends on the metric tensor $g_{\mu\nu}$ and, perhaps, also on 
some extra-matter field(s)
$\psi$. For geometric scalarisation the latter are not necessary. 
The corresponding equation of motion for the scalar field $\phi$
reads
\begin{eqnarray}
\label{eq-phi}
\Box \phi=\frac{\partial f}{\partial \phi}{\cal I} \ .
\end{eqnarray}

The occurrence of spontaneous scalarisation requires two different ingredients. Firstly, there is a \textit{scalar-free solution} of (\ref{eq-phi}) with
\begin{eqnarray}
 \phi=\phi_0 \ ,
\end{eqnarray}
everywhere. This demands the coupling function should satisfy the condition
\begin{eqnarray}
\label{condx}
\frac{\partial f}{\partial \phi}\Big |_{\phi=\phi_0}=0 \ .
\end{eqnarray}
One can set $\phi_0=0$ without any loss of generality (via a field redefinition).
Thus, the usual vacuum (or electro-vacuum) BHs of general relativity also solve the considered model. 
Secondly, the model should possess another set of solutions with a nontrivial scalar field -- {\it the scalarised (or hairy) BHs}.
These solutions are usually entropically preferred
over the scalar-free ones  ($i.e.$ they maximize the entropy for given global charges).
Moreover, they are smoothly connected with the scalar-free set, approaching it for $\phi=0$.

At the linear level, the spontaneous scalarisation phenomenon manifests itself as a tachyonic instability when scalar perturbations of the scalar-free solution are studied. For this analysis one considers a small-$\phi$, denoted $\delta\phi$, expansion of the coupling function
\begin{eqnarray}
\label{f-phi-small}
f(\phi)=f|_{\phi=0}+ \frac{1}{2} \frac{\partial^2 f}{\partial \phi^2}\Big |_{\phi=0}  \delta \phi^2+\mathcal{O}(\delta \phi^3) \ .
\end{eqnarray}
Then, the linearised form of equation (\ref{eq-phi}) reads
\begin{eqnarray}
\label{eq-phi-small}
(\Box-\mu_{\rm eff}^2)\delta \phi =0\ , \qquad {\rm where}~~ \mu_{\rm eff}^2\equiv  \frac{\partial^2 f}{\partial \phi^2}\Big |_{\phi=0} {\cal I} \ .
\end{eqnarray}
A tachyonic mass $\mu_{\rm eff}^2<0$ signals an instability of the scalar-free solution; $\mu_{\rm eff}^2<0$ is also the condition for the existence of bound state solutions of eq. (\ref{eq-phi-small}). Such bound states mark the onset of the instability. The tachyonic condition can be satisfied for suitable choices of  the source ${\cal I} $.

The two types of scalarisation discussed in the Introduction depend on the `source' term ${\cal I}$. Geometric scalarisation has been considered using the Gauss-Bonnet invariant $\mathcal{L}_{GB}$ as the source term in~\cite{Silva:2017uqg,Doneva:2017bvd,Antoniou:2017acq}
\begin{eqnarray}
 {\cal I}=\mathcal{L}_{GB} \ .
\end{eqnarray}
Similar solutions should exist when taking instead a source term
given by the 2nd four dimensional topological invariant
\begin{eqnarray}
 {\cal I}=\mathcal{L}_{CS} \ ,
\end{eqnarray}
where $\mathcal{L}_{CS}$ is the Pontryagin density, 
as shown in~\cite{Brihaye:2018bgc}.
The latter, however, requires the presence of rotation for a vacuum BH to become scalarised.

Matter-induced scalarisation is illustrated by the recent work~\cite{Herdeiro:2018wub},
which studied the spontaneous scalarisation of electrovacuum BHs, and where the source term was
\begin{eqnarray}
 {\cal I}=F_{\mu \nu}F^{\mu \nu} \ .
\end{eqnarray}
In this case, the precise form of the coupling function $f(\phi)$ does not seem to be important. The concrete results in 
\cite{Herdeiro:2018wub} were found
for
$
 f(\phi)=e^{-\alpha \phi^2},
$
where the coupling constant $\alpha$ is an input parameter.
%
%

\subsection{Scalarisation due to the $\phi^2 R$ non-minimal coupling}
In this work we shall consider matter-induced scalarisation with 
 \begin{eqnarray}
\label{our-model}
f(\phi)= \frac{1}{2}\xi \phi^2 \ , \qquad {\rm and} \qquad {\cal I}=R \ ,
\end{eqnarray}
$i.e.$ a scalar field Lagrangian 
 \begin{eqnarray}
\label{L-phi}
{\cal L}_\phi=  - \frac{1}{2}(\nabla \phi)^2 - \frac{1}{2}\xi \phi^2 R \ ,
\end{eqnarray}
where $\xi$ is a dimensionless coupling constant, which is an input parameter of the theory.

The non-minimal coupling $\xi \phi^2 R$ has a long history starting with~\cite{Chernikov:1968zm,Callan:1970ze} - see also the review discussions in~\cite{Capozziello:2011et,Faraoni:2000gx,Faraoni:1998qx}. 
Essentially, a non-minimal coupling $\xi\neq 0$ is sourced by quantum corrections: even if $\xi$ is set to zero 
in the classical action, renormalisation makes $\xi\neq 0$, see also~\cite{Birrell:1982ix}.
Thus, $\xi$=0, $i.e.$, ``minimal coupling" is a classical value; amongst the non-zero values of $\xi$,   ``conformal coupling"
($\xi$=1/6) corresponds to the case for which a massless scalar field theory becomes conformally invariant (in four spacetime dimensions)~\cite{Birrell:1982ix}. 

The literature on this sort of non-minimal coupling and its physical implications, in particular for cosmology, is vast and we do not intend to review it here. In the context of compact objects, nonetheless, we would like to mention the Bronnikov-Melnikov-Bacharova-Bekenstein (BMBB) BH with conformal scalar hair\footnote{We remark that  the
BMBM BH is rather special. The spherically symmetric asymptotically flat BHs 
with generic $\xi$ cannot support non-minimally coupled spatially regular
neutral scalar fields~\cite{Mayo:1996mv, Bekenstein:1996pn,Hod:2017ssh,Hod:2017hvl}.}
\cite{Bekenstein:1974sf,BBM}, traversible wormholes
\cite{Barcelo:1999hq,Barcelo:2000zf},  as well as  solitons and BHs with a non-minimally coupled gauged Higgs field~\cite{vanderBij:2000cu,Nguyen:1993ep,Brihaye:2014vba}, see also~\cite{Faraoni:1998qx,Capozziello:2011et}.

Here, we shall investigate under which circumstances BH scalarisation occurs for the simple model~(\ref{our-model}). From the examples studied below, our main conclusion, that we conjecture holds generically, is that given a (static, spherically symmetric, asymptotically flat)
 BH with a nonvanishing Ricci scalar, one always finds scalarised generalisations, regular on and outside an event horizon and asymptotically flat, for  
(infinite) discrete set of $\xi$.

Three concrete of BHs with non-vanishing Ricci scalar shall be considered below, of which we shall construct their scalarised counterparts. Firstly we consider the Einstein-Maxwell-dilaton BH obtained by Garfinkle, Horowitz and Strominger (GHS)~\cite{Garfinkle:1990qj}. We regard this as an example of non-conformally invariant classical matter, as this model emerges in string theory at tree level. Secondly, we consider an Einstein-Maxwell model with $F^4$ correction. This can be faced in the same spirit of the Euler-Heisenberg effective Lagrangian in QED, $\mathcal{L}_{\rm EH} \propto F^2+ a(F^2)^2+b(F\star F)^2$,  which accounts for vacuum polarisation, but omitting the $(F^2)^2$ term. This simplification allows for a simple closed form BH solutions which we present here for the first time.  We do not expect that including the full Euler-Heisenberg Lagrangian, however, should change qualitatively our conclusions concerning scalarisation. The final example considers the scalarisation of the NCG inspired generalisation of the Schwarzschild BH proposed by Nicolini, Smailagic and Spallucci (NSS)~\cite{Nicolini:2005vd}.  

In all these examples we start by constructing the corresponding zero modes, or \textit{scalar clouds}, $i.e.$ an infinitesimally small scalar field bound states.
For a given $\xi$ these occur for specific background parameters. The continuation at the non-linear level of these scalar clouds yields the scalarised BHs.
For studies of a similar scalarisation in the context of horizonless configurations see, $e.g.$,~\cite{Salgado:1998sg,Fuzfa:2013yba}.

\section{Setup}
\label{sec3}

Consider the model described by the action
\begin{eqnarray}
\label{action}
\mathcal{S}=\int d^4 x \sqrt{-g}
\left[ 
\frac{R}{4}+{\cal L}_0 (\Psi)
\right] +\mathcal{S}_\phi\ ,
\end{eqnarray}
where $\mathcal{L}_0$ is the Lagrangian for the matter fields, collectively denoted by $\Psi$, and the scalar field action is provided by~(\ref{actionS}) with (\ref{our-model}), for the case of interest herein. Variation of action $w.r.t.$ the metric tensor leads to the Einstein equations
\begin{eqnarray}
\label{Einstein-eqs}
R_{\mu \nu}-\frac{1}{2}Rg_{\mu \nu}=2 (T_{\mu\nu}+T_{\mu\nu}^\phi) \ ,
\end{eqnarray}
where  
$T_{\mu\nu}$
is the energy momentum tensor associated with ${\cal L}_0$ and $T_{\mu\nu}^{(\phi)}$ is the one associated to the scalar field action 
$\mathcal{S}_\phi$.
The equations of motion for the matter fields can be written once the respective Lagrangian is specified. We assume the existence of a scalar-free BH solution,  with $\phi=0$, which solves the scalar field equation
\begin{eqnarray}
\label{KG-eq}
 \nabla^2\phi-\xi R \phi=0 \ .
\end{eqnarray}
$R\neq 0$ allows, in principle, to circumvent the standard Bekenstein-type~\cite{Bekenstein:1972ny} no-scalar hair theorems. This type of argument is based on constructing an identity that implies triviality of the scalar field. For instance, restricting to static configurations,
one constructs from (\ref{KG-eq}) the identity
\begin{eqnarray}
\label{id1}
\int d^3 x \sqrt{-g}
\left[
(\nabla \phi)^2+\xi R\phi^2
\right]=0 \ .
\end{eqnarray}
The kinetic term is everywhere non-negative, but it is clear that if $\xi R<0$ for some space region,
 relation (\ref{KG-eq}) $cannot$ be used to exclude the existence of solutions.

In the following, we shall be considering the generic spherically symmetric line element
\begin{eqnarray}
\label{gen-metric}
ds^2=-\sigma^2(r)N(r)dt^2+\frac{dr^2}{N(r)}+P^2(r)(d\theta^2+\sin^2\theta d\varphi^2) \ ,
\end{eqnarray}
where $\sigma,N,P$ are radial functions to be determined. Observe that we have kept some metric gauge freedom, which shall be conveniently fixed later. For future reference, the corresponding Ricci scalar is
\begin{eqnarray}
\label{Rgen}
R=-N''-N'\left(\frac{4P'}{P}+\frac{3\sigma'}{\sigma}\right)
-2N\left(\frac{2P''}{P}+\frac{\sigma''}{\sigma}+\frac{2P'\sigma'}{P\sigma}\right)
+\frac{2}{P^2}(1-NP'^2) \ ,
\end{eqnarray}
where the prime denotes radial derivative. 
%

\subsection{Scalar clouds}

Before studying the fully non-linear problem it is instructive
to consider  the limit wherein the scalar field $\phi$ is infinitesimally small, denoted as $\delta \phi$ on the background of the scalar-free BH solution we assume to exist. Thus, we only have to solve the scalar field equation (\ref{KG-eq}), on the fixed background (\ref{gen-metric}) which solves the Einstein-$\Psi$ equations.

Taking the usual multipolar decomposition 
\begin{eqnarray}
\label{p1}
 \delta\phi=Y_{\ell m}(\theta,\varphi)U_\ell(r) \ ,
\end{eqnarray}
where $Y_{\ell m}$ are the real spherical harmonics and $\ell,m$
are the associated quantum numbers, $i.e.$ $\ell=0,1,\dots$ and $-\ell\leqslant m \leqslant \ell$,
one finds 
the following radial equation for the function $U_\ell$ 
\begin{eqnarray}
\label{zero-general}
\frac{1}{P^2\sigma}(P^2N\sigma U_\ell')'-\left[ \frac{\ell(\ell+1)}{P^2} +\xi R \right] U_\ell=0 \ ,
\end{eqnarray}
where $R$ is given by (\ref{Rgen}).

We are interested in test scalar field configurations around an asymptotically flat, scalar-free, BH background. At the (non-extremal) BH horizon, located at $r=r_h$, the metric functions are assumed to have a generic power series expansion of the form
\begin{equation}
N(r)=N_1(r-r_h)+\dots \ ,    \ P(r)=P_h+P_1(r-r_h)+\dots \ ,  \   \sigma(r)=\sigma_h+\sigma_1(r-r_h)+\dots \ . 
\label{expansions}
\end{equation}
The test scalar field configurations we seek (scalar clouds, or zero modes) are regular at the horizon and vanish asymptotically, being smooth everywhere. At the horizon, the radial function describing the scalar field reads:
\begin{eqnarray}
\label{ss1}
U_\ell(r) =\phi_0+\phi_1(r-r_h)+\mathcal{O}(r-r_h)^2 \ , 
\end{eqnarray}
where, from the scalar field equation, 
\begin{eqnarray}
\label{ss2}
 \phi_1=  \frac{1}{N_1 P_h^2}
\big[
\ell(\ell+1)+\xi P_h^2 R(r_h) 
\big] \phi_0 \ .
\end{eqnarray}
At infinity, on the other hand,  asymptotic flatness of the BH background implies 
\begin{equation}
N= 1-\frac{2M}{r}+\dots\ , \qquad P \to r \ , \qquad  \sigma\to 1 \ ,
\label{expansioninf}
\end{equation} 
where $M$ is the BH mass.  For the scalar field,  one finds
\begin{eqnarray}
\label{sinf}
U_\ell(r) = \frac{Q_s}{r^{\ell+1}}+  \dots \ ,
\end{eqnarray} 
where $Q_s$ is a constant. This constant, as well as the scalar field solution interpolating between the horizon and infinity, is found by solving numerically equation (\ref{zero-general}).
This corresponds to an eignevalue problem, which for a given 
background (and a number $\ell$), selects an infinite set of coupling constants
$\xi_n=\{ \xi_0, \xi_1,\dots \}$
labeled by the number $n$ of nodes of the scalar amplitude $U_\ell$.

\subsection{The scalarised BHs}
The scalarised BH solutions are the non-linear continuation of these `scalar clouds'  which solve the full model (\ref{action}) with non-trivial $\phi$.
The scalar field energy-momentum tensor in~(\ref{Einstein-eqs}) is
\begin{eqnarray}
\label{Tiks}
&&
T_{\mu \nu}^{(\phi)}=
 \phi_{ , \mu}\phi_{,\nu}
-\frac{1}{2}g_{\mu\nu}   
g^{\alpha \beta} 
 \phi_{ , \alpha }\phi_{,\beta} 
 + \xi
\left[
                                              \left(
R_{\mu \nu}-\frac{1}{2}g_{\mu\nu}R 
                                             \right)\phi^2
+g_{\mu\nu}\Box \phi^2
-\phi^2_{;\mu\nu}
\right] \ .
\end{eqnarray}  
No other explicit coupling between the scalar field  
$\phi$ and the other matter fields $\Psi$ will be considered below.

Only the $\ell=0$ (test field) mode leads to a spherically symmetric solution, 
compatible with the line-element (\ref{gen-metric}). Thus, these are the modes we shall focus in here. Moreover, all scalarised BH solutions  in this work are found for the metric gauge choice
\begin{eqnarray}
\label{gc}
P(r)=r\ , \qquad {\rm and} \ \ N(r)=1-\frac{2m(r)}{r} \ ,
\end{eqnarray}
where $m(r)$ is the Misner-Sharp mass function.
With this setup, the  equations for the matter fields $\Psi$
are solved together with the Einstein equations for $N(r),\sigma(r)$
and the scalar field, $\phi=U_0(r)$.  This results in a standard boundary value problem.
As mentioned before, the BH horizon is located at $r=r_h>0$ and the expansions (\ref{expansions})-(\ref{ss1}) hold to leading order, introducing the positive constants $N_1,\sigma_h,\phi_0$.
At infinity, the expansions~(\ref{expansioninf})-(\ref{sinf}) hold, introducing another two constants: $M$, the ADM mass and $Q_s$, the scalar 'charge'.
The asymptotic behaviour of the
matter fields $\Psi$ 
is  similar to that in the $\phi=0$ case.
Finding a solution of the problem compatible with these asymptotics
requires a fine-tuning of the data at the horizon, as specified $e.g.$ by the value of the scalar field.

For a given value of $\xi$, which is an input parameter of the theory,
solving this boundary value problem results in branches of scalarised BHs,
which are labeled by an integer $n$, describing the number of nodes of scalar field $\phi$.
Only nodeless solutions ($n=0$) are reported in this work.
Furthermore, to simplify the analysis, we shall restrict ourselves herein to the analysis of non-extremal BHs.
Finally, we shall not attempt to clarify the  critical behaviour of solutions.

A particularly relevant quantity in the following is the BH entropy, $S$, of the scalarised  BHs. 
In the absence of a supplementary contribution from ${\cal L}_0 (\Psi)$ in (\ref{action}),
the expression of $S$,
as derived by using Wald's formalism
\cite{Wald:1993nt,Iyer:1994ys},
possesses
an extra contribution with respect to that in Einstein's gravity, 
due to the non-minimal coupling with the scalar field, and reads 
\begin{eqnarray}
\label{S}
S=\pi r_h^2(1-2\xi \phi_0^2)\ .
\end{eqnarray}
Following standard conventions, we also define the reduced area $a_H$, temperature $t_H$, entropy $s$ and charge $q$:
\begin{eqnarray}
\label{scale1}
a_H\equiv \frac{A_H}{16\pi M^2}\ , \qquad t_H\equiv 8\pi T_H M\ , \qquad s\equiv \frac{S}{4\pi M^2}\ , \qquad q\equiv\frac{Q}{M} \ ,
\end{eqnarray}
where $A_H,T_H,Q$ are the BH area, temperature and 
electric charge. 
The latter will only be present for the cases with a Maxwell field below and it will be slightly modified in Section~\ref{sec5}.

\section{Scalarised dilatonic BHs}
\label{sec4}

\subsection{The scalar-free solution}
Our first example is purely classical.  We consider the GHS solution~\cite{Garfinkle:1990qj}, which provides a simple example of a classical  
BH beyond electro-vacuum that has a non-vanishing Ricci scalar, being a solution of low energy string theory (for a particular choice of the parameter $a$ introduced below). More concretely, this family of BH solutions solves Einstein's gravity coupled to a Maxwell field $F=dA$ and a dilaton $\psi$. The corresponding matter action term  in (\ref{action})
reads
\begin{eqnarray}   
{\cal L}_0(A,\psi)=  -\frac{1}{2}(\nabla \psi)^2 
-
\frac{1}{4}e^{2 a\psi}F^2 \ ,
\end{eqnarray}
where $a$ is a free parameter which governs the strength of the coupling between the dilaton and
the Maxwell field.\footnote{
Changing the sign of $a$ is
equivalent to changing the sign of $\psi$; it thus suffices to consider $a\geqslant 0$.}
When $a=0$, the action reduces to the usual Einstein-Maxwell
theory with a decoupled massless real scalar. When $a=1$, this model is part of the low energy action of string theory; 
$a=\sqrt{3}$ corresponds to the Kaluza-Klein value, that is the Einstein-Maxwell-dilaton theory that emerges from the dimensional reduction of vacuum gravity in five spacetime dimensions.

The GHS BH solution has the line element~(\ref{gen-metric}) with 
\begin{equation}  
\label{GHS-solution}  
\sigma(r)=1\ , \qquad 
N(r)
=\left(1-\frac{r_+}{r} \right)\left(1-\frac{r_-}{r}\right)^{\frac{1-a^2}{1+a^2}}\ , \qquad
P=r \left(1-\frac{r_-}{r}\right)^{\frac{ a^2}{1+a^2}}\ , 
\end{equation}
together with the Maxwell potential and dilaton field
\begin{equation}
A=\frac{Q}{r}dt \ , \qquad 
e^{2\psi}=\left(1-\frac{r_-}{r}\right)^{\frac{2a}{1+a^2}}\ .
\end{equation}
The two free parameters $r_{+}$, $r_{-}$ 
(with $r_-<r_+$)
 are related to the
ADM mass, $M$, and (total) electric charge, $Q$, by
\begin{eqnarray}   
  M = \frac{1}{2}
	\left[
	r_+ +\left(\frac{1-a^2}{1+a^2}\right)r_-
	\right] \ , \qquad 
	Q=\left(  
	\frac{r_-r_+}{1+a^2}
	\right)^{\frac{1}{2}} \ .
\end{eqnarray}
For all $a$, the
surface $r= r_+$ is the location of the event horizon.
In the extremal limit, which corresponds to the coincidence limit $r_- = r_+$, 
 the area of the event horizon goes to zero for $a\neq 0$.
The Hawking temperature, however, only goes to zero in the extremal limit for $a<1$,
while for $a=1$ it
approaches a constant, and for $a>1$ it diverges.
The Ricci scalar of the GHS BHs is strictly positive outside the horizon,
 \begin{eqnarray}   
R=\frac{2a^2 r_-^2}{(1+a^2)r^4}\left(1-\frac{r_+}{r}\right)\left(1-\frac{r_-}{r}\right)^{-\frac{1+3a^2}{1+a^2}}>0 \ ,
\end{eqnarray}
which implies that scalarization can occur for negative $\xi$ only. 
In what follows, we focus on the case $a=1$, although we have verified the  occurrence of 
scalarisation for several other values of $a$ as well.

\subsection{The scalarised solutions}

Starting with the analysis of zero modes,
we have found that, for any fixed $\ell$, equation 
(\ref{zero-general})
solved on the GHS background with given $M,Q$
admits
a family of non-trivial solutions (with $\phi(\infty)=0$) for a discrete spectrum
of values of the coupling parameter
$\xi_n$, which are labelled by the node number $n$, although only nodeless solutions $n=0$ are considered here (and $\xi_0$ is denoted $\xi$ for simplicity). This feature occurs for any value of the global charges $(M,Q)$. 

\begin{figure}[h!]
\begin{center}
\includegraphics[width=0.55\textwidth]{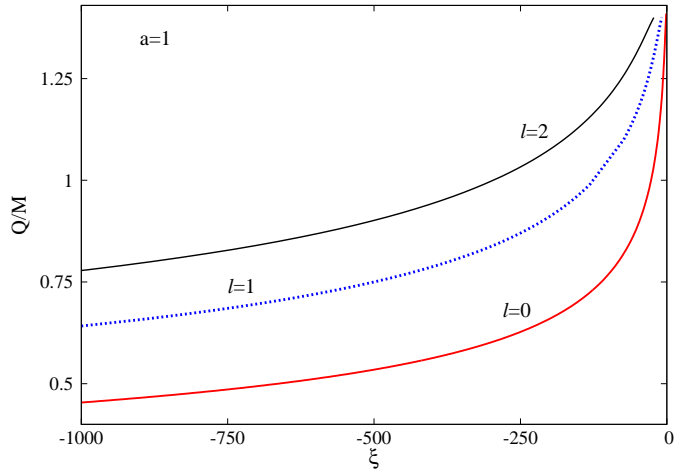}  
\caption{ 
$\ell=0,1,2$ existence lines for the GHS
solutions with $a=1$, plotted for $q\equiv Q/M$ in terms of the coupling $\xi$. Scalarisation only occurs for negative $\xi$. A larger coupling (in modulus) implies the scalarisation instability becomes possible for BHs with smaller $q$.
}
\label{existence-lines-GHS}
\end{center}
\end{figure}

In Fig. \ref{existence-lines-GHS} we exhibit the value of $q$ for which the zero mode occurs, as a function of the coupling constant $\xi$. This defines a line - \textit{existence line} - which is exhibited for the nodeless zero modes and $\ell=0,1,2$.  Recall that this zero mode is at the onset of the instability of the scalar-free BH. That is, BHs with a larger (smaller) $q$ than that at which the zero mode occurs (for fixed $\xi,\ell$) are unstable (stable) against scalarisation due to that $\ell$-mode. Fig. \ref{existence-lines-GHS} shows that the onset of the scalarisation instability occurs for BHs with increasingly smaller charge to mass ratio as one increases $|\xi|$. This resembles the pattern observed in~\cite{Herdeiro:2018wub}. We recall that for the GHS solutions $q$ can exceed unity; in fact $q=\sqrt{2r_-/r_+}$ and thus it has a limiting value of $q=\sqrt{2}$.  Only solutions with large $q$ (and greater than unity) can become scalarised for small $|\xi|$, and for each $\ell$ there is a minimum value of $|\xi|$ for which scalarisation occurs. 
The corresponding (approximate) values for $\ell=(0,1,2)$ are, respectively, $(-1.124,-8.877,-22.291)$. In this limit the horizon area goes to zero and $q\rightarrow \sqrt{2}$.

\begin{figure}[h!]
\begin{center}
{\includegraphics[width=9cm]{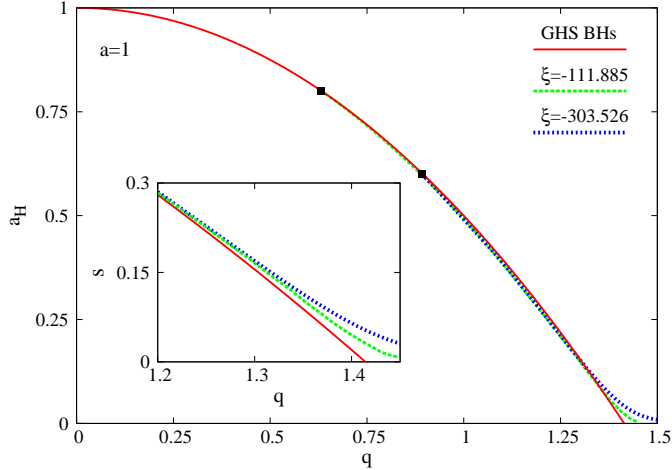}}
\caption{ 
(Main panel) reduced area $vs.$ $q$ for GHS BHs and the corresponding scalarised BHs. The former fall on the solid red line; the latter on the blue and green dashed lines for two values of the coupling. (Inset) Reduced entropy $vs.$ $q$ for the same solutions. The shown $q$ interval contains scalarised BHs that have a smaller reduced area than the scalar-free ones, but the reduced entropy is always larger for the former.  
}
\label{bifurcationGHS}
\end{center}
\end{figure}

As expected, these scalar zero modes can be continued into the fully non-linear regime yielding scalarised BHs. For this construction we have only analysed the spherically symmetric solutions, which bifurcate from the scalar cloud with $\ell=0$, as mentioned before. The bifurcation points, for two different values of $\xi$, together with the corresponding line of scalarised BHs can be seen in Fig.~\ref{bifurcationGHS}, in a reduced area $vs.$ $q$ diagram. An interesting feature is that the reduced area of the scalarised BHs can be smaller than that of the comparable scalar-free BH (with the same $q$) but the reduced entropy, seen in the inset of the same figure is always larger, showing the scalarised BHs are entropically favoured. The latter conclusion is general for all solutions analysed. 

\begin{figure}[h!]
\begin{center}
\includegraphics[width=0.495\textwidth]{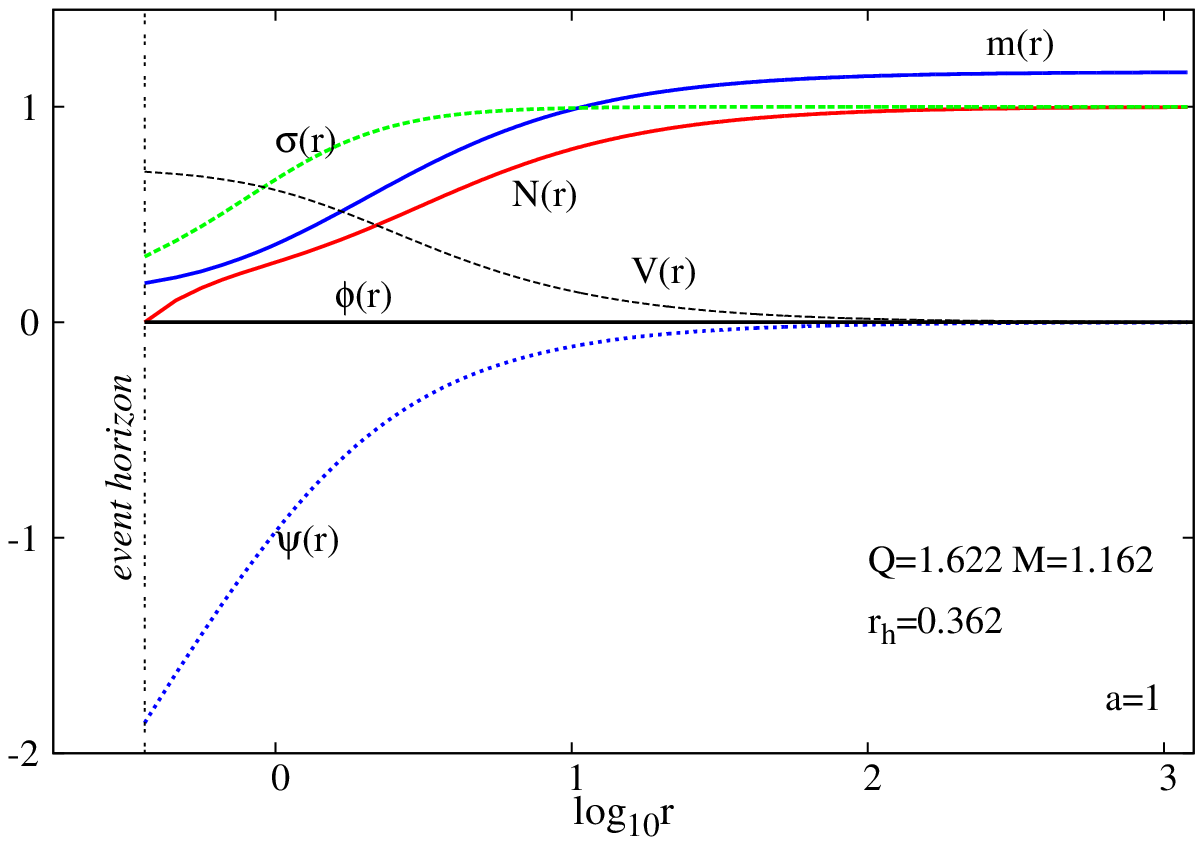} 
\includegraphics[width=0.495\textwidth]{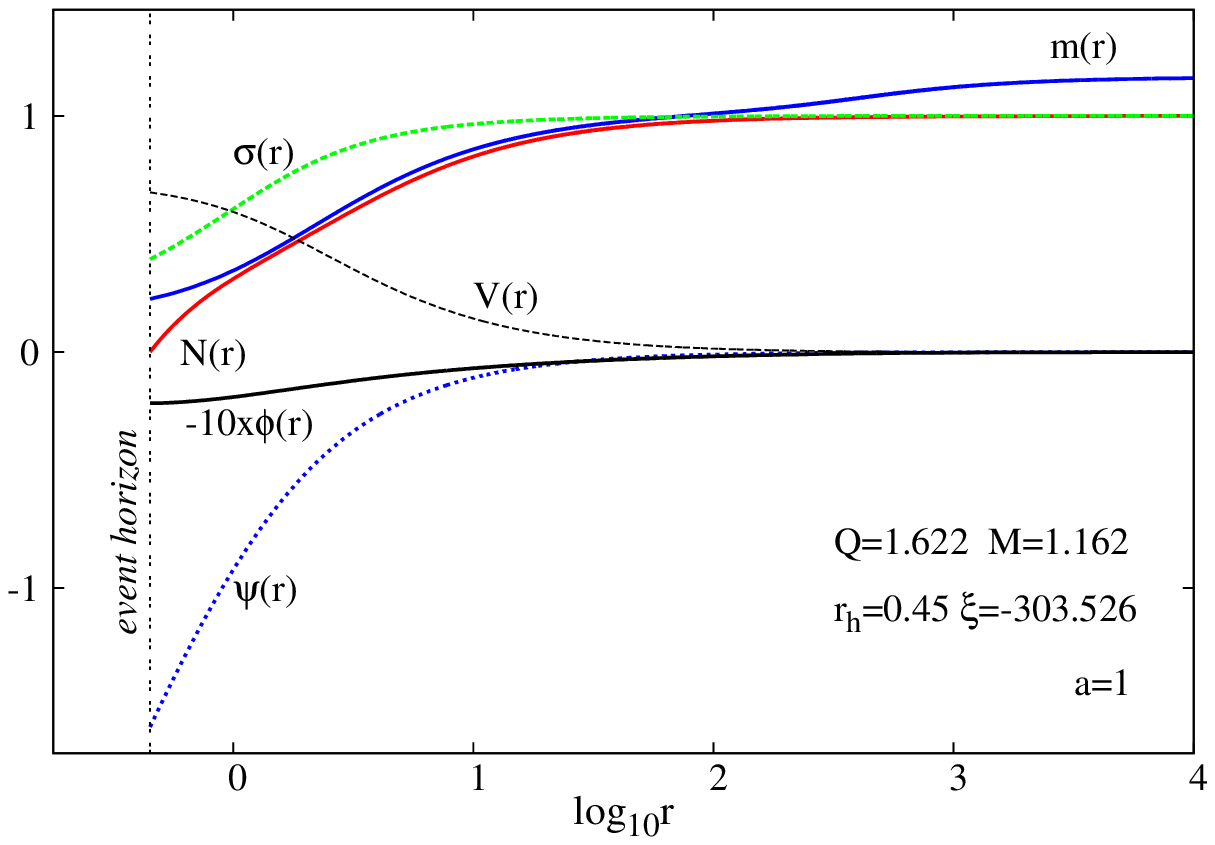}  
\caption{ 
The profile functions of a typical GHS solution (left panel) and its scalarised version (right panel) for the same global charges.
}
\label{profileGHS}
\end{center}
\end{figure}

\begin{figure}[ht!]
\begin{center}
{\includegraphics[width=7.8cm]{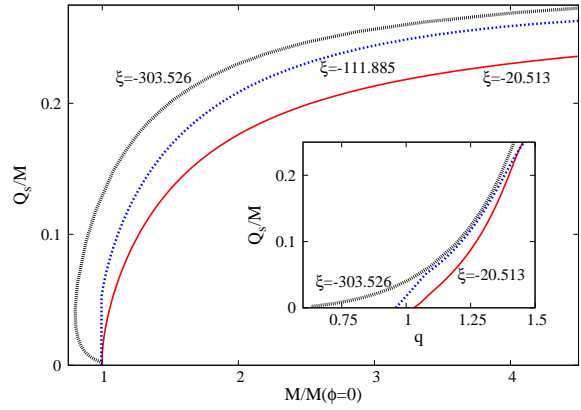}}
\caption{
Variation of the reduced scalar charge with: (main panel) the ratio of the ADM mass to the mass of the scalarised solution at the bifurcation point; (inset) the electric charge to mass ratio, $q$.  
}
\label{sGHS}
\end{center}
\end{figure}

The profile functions describing the line element, dilaton and scalar fields, and electric potential $V(r)$ of a typical scalarised solution is shown in Fig.~\ref{profileGHS}, comparing it with those for the GHS scalar free solution with the same global charges. The differences are visible although of small magnitude. The variation of the scalar charge for three values of the coupling, in terms of the total mass normalised to the mass of the bifuraction point scalar-free solution and also in terms of the charge to mass ratio, are shown in Fig.~\ref{sGHS}. Finally, we remark that, similarly to the $\phi=0$ case, these solutions seems to possess an extremal limit with vanishing horizon area. 

\section{Scalarised Reissner-Nordstr\"om-$F^4$  BHs}
\label{sec5}
Our second example pertains a class of quantum corrected BHs where the quantum correction emerges from the matter sector. It is well known that classical electrodynamics loses conformal invariance due to vacuum polarisation. The Euler-Heisenberg effective field theory~\cite{Heisenberg:1935qt} takes into account the scattering of photons by virtual electron-positron pairs. Integrating the corresponding loop produces an effective quartic interaction vertex between photons, thus a non-linear theory of electrodynamics which is not scale invariant. Another relevant non-linear theory of electrodynamics emerging from quantum corrections is Born-Infeld theory, in the context of the open string~\cite{Fradkin:1985qd}. In the following we shall first remark that, generically, non-linear electrodynamics breaks down the scale invariance of Maxwell's theory, before presenting a novel BH solution in this context, that we dub RN-$F^4$ BH, that will be analysed below in the context of scalarisation.

\subsection{Scale invariance breakdown in non-linear electrodynamics}

Following the discussion in 
\cite{Gibbons:2000xe},
we consider a general Lagrangian ${\cal L} = {\cal L}(x, y)$ depending on the only two Lorentz and gauge invariant scalars that can be constructed for the electromagnetic field in four spacetime dimensions
\begin{eqnarray}
x\equiv  \frac{1}{4}F_{\mu \nu}F^{\mu \nu} \ , \qquad {\rm and} \qquad y \equiv \frac{1}{4}F_{\mu \nu}(\star F)^{\mu \nu} \ , 
\end{eqnarray}
where we have defined the dual field strength as
\begin{eqnarray}
 (\star F)_{\mu\nu }=\frac{1}{2}\epsilon_{\mu\nu \alpha \beta  }F^{\alpha \beta} \ ,
\end{eqnarray}
and $\epsilon_{\mu\nu \alpha \beta  }$ is the Levi-Civita tensor. For such a generic Lagrangian, the corresponding energy-momentum tensor reads
\begin{eqnarray}
\label{genericT}
T_{\mu \nu}={\cal L}_{,x} T_{\mu \nu}^{(Maxwell)}+\frac{T}{4}g_{\mu \nu} \ ,
\end{eqnarray}
where the comma denotes partial derivative, the (usual) Maxwell energy-momentum tensor reads
\begin{eqnarray}
T_{\mu \nu}^{(Maxwell)}=F_{\mu\alpha}  F_{\nu\beta}g^{\alpha \beta} -\frac{1}{4}g_{\mu\nu }F_{\alpha\beta}F^{\alpha\beta} \ ,
\end{eqnarray}
and the trace is
\begin{eqnarray}
T\equiv T_\mu^\mu=-4({\cal L}-x {\cal L}_{,x}-y {\cal L}_{,y}) \ .
\end{eqnarray} 
One can easily see that
the condition $T=0$ is violated in a generic non-linear Maxwell theory.  
It follows that the corrected RN BHs emerging in electro-vacuum plus non-linear electrodynamics corrections will, in general, possess a non-vanishing Ricci scalar
$R=-T \neq 0$.

\subsection{The scalar-free solution: a new exact BH}

A minimal deviation from standard Maxwell action - which corresponds to $\mathcal{L}=x$ - requiring both magnetic and electric fields,
is found by adding a quadratic term  in $y$ and reads\footnote{We recall that 
$F_{\mu \nu}\star F^{\mu \nu}$
is a total derivative.
As such, a linear correction in $y$ to ${\cal L}$ leads to the same  solutions as in the Maxwell case.}
\begin{eqnarray}
\label{F2F4}
 {\cal L}_{0}(A)=x+\alpha y^2
= \frac{1}{4}F^2 + \frac{ \alpha}{16} F^4\ , \qquad {\rm with}~~F^4\equiv [F_{\mu \nu}(\star F)^{\mu \nu}]^2 \ ,
\end{eqnarray}
where $\alpha$ is a dimensionful  constant, which is a new parameter of the theory. 
It plays a similar role, say,  to the $\alpha$ coupling constant in front of the GB term in the Einstein-GB model.
As further motivations to the model (\ref{F2F4}), we mention that the quartic term occurs, say, in Born-Infeld theory \cite{Tseytlin:1997csa}
or in the higher loop corrections to the $d = 10$ heterotic string low energy effective action. Further motivation can be found in Sec. VII of Ref. \cite{Maleknejad:2011sq}. The BH solutions of a similar model with non-Abelian gauge fields have been studied in~\cite{Radu:2011ip,Herdeiro:2017oxy}, and possess a variety of interesting solution ($e.g.$ the RN solution possesses non-Abelian hair).
We remark that the requirement of a positive energy density imposes $\alpha<0$ in (\ref{F2F4}),  which is the only case considered in this work.

Model (\ref{F2F4}) has an energy-momentum tensor with a non-vanishing
trace $T=4\alpha y^2$. Thus, the corresponding BH solutions 
 possess a nonzero Ricci scalar and are subject to scalarisation in the context of the total action~(\ref{action}).
Moreover, we shall now present a new (scalar-free) exact dyonic BH solution which contains some interesting features and that can get scalarised in the context of the present paper. 

To find this corrected RN BH,
we solve the Einstein equations (\ref{Einstein-eqs})
with
an energy-momentum tensor 
which can be read off from  (\ref{genericT}).
The equations of motion also include the gauge field equations which read
\begin{eqnarray}
\label{M-eqs}
\nabla_\mu F^{\mu \nu}+\frac{1}{2} \alpha (\star F)^{\mu \nu}[F_{\alpha\beta}(\star F)^{\alpha\beta}]_{,\mu}=0 \ .
\end{eqnarray}

We choose again the metric ansatz (\ref{gen-metric}) and gauge choices (\ref{gc}), 
and consider  a 
gauge connection which contains both an electric and magnetic part:
\begin{eqnarray}
A=V(r) dt +Q_m \cos \theta d\varphi\ ,
\end{eqnarray}
$V(r)$ being the electric potential and 
$Q_m$ the magnetic charge. Then, it is straighforward to construct an
exact solution of the equations (\ref{Einstein-eqs}),  (\ref{M-eqs}).
The corresponding expression simplify by defining the new length scale $r_0$
\begin{eqnarray} 
r_0\equiv (-128 \alpha Q_m^2 )^{1/4} \ .
\end{eqnarray}
Then, the solutions - the RN-$F^4$ BHs - are defined by the profile functions
\begin{eqnarray}
\label{newF4}
\sigma(r)=1\ , \qquad 
N(r)=1-\frac{2M}{r}+\frac{Q_m^2+Q^2}{r^2}+\frac{Q}{r}\Psi(r)\ , \qquad V(r)=\frac{Q}{r}+\Psi(r) \ ,
\end{eqnarray}
and
\begin{eqnarray} 
\Psi(r)\equiv -\frac{Q}{r}
\left\{
1+\frac{r}{4r_0} 
               \left[
			\arctan  \left (\frac{\frac{r}{r_0}}{1-\frac{r^2}{2r_0^2} } \right)
			+\frac{1}{2}
			\log \left( \frac{\frac{r^2}{2r_0^2}-\frac{r}{r_0}+1}{\frac{r^2}{2r_0^2}+\frac{r}{r_0}+1} \right)
							 \right] 
\right\} \ .
\label{f4psi}
\end{eqnarray}
As for the standard  RN BH, $M, Q$ are constants, corresponding to 
ADM mass and total electric charge of the solution, respectively.

The position of the (outer) Killing horizon, $i.e.$ the event horizon,  
is determined as the largest (positive) root of the equation
$N(r_h)=0$, as for the standard RN BH.
Unlike the latter, however, $r_h$ cannot be expressed as a function of  $M,Q,Q_m$; it is implicitly defined by the equation:
\begin{equation} 
M=\frac{r_h}{2}
+\frac{Q^2+Q_m^2}{2r_h}-\frac{Q^2}{2r_h}
\left\{
1+\frac{r_h}{4r_0} 
               \left[
			\arctan  \left (\frac{\frac{r_h}{r_0}}{1-\frac{r_h^2}{2r_0^2} } \right)
			+\frac{1}{2}
			\log \left( \frac{\frac{r_h^2}{2r_0^2}-\frac{r_h}{r_0}+1}{\frac{r_h^2}{2r_0^2}+\frac{r_h}{r_0}+1} \right)
							 \right]
\right\} \ .
\end{equation}
The Hawking temperature and area of a spatial section of the event horizon of the solution are 
\begin{eqnarray} 
T_H=\frac{1}{4\pi r_h}
\left[
1-\left(\frac{ Q_m^2}{ r_h^2}+\frac{Q^2 r_h^2}{4r_0^4+r_h^4}\right)\right]\ , \qquad A_H=4\pi r_h^2 \ .
\end{eqnarray}
One can verify that the solutions satisfy the first law of BH thermodynamics in the form
 \begin{eqnarray} 
dM=T_H\frac{1}{4}dA_H+\Phi_e dQ+\Phi_m dQ_m \ ,
\end{eqnarray}
with the chemical potentials
\begin{eqnarray} 
\Phi_e=-\frac{Q}{4r_0}
\left[
\arctan           \left(
           \frac{\frac{r_h}{r_0}}{1-\frac{r_h^2}{2r_0^2} } \right)
			+\frac{1}{2}
			\log \left( \frac{\frac{r_h^2}{2r_0^2}-\frac{r_h}{r_0}+1}{\frac{r_h^2}{2r_0^2}+\frac{r_h}{r_0}+1} 
			            \right)
\right]\ , \qquad \Phi_m=\frac{Q_m}{r_h} \ .
\end{eqnarray}

The reduced quantities 
$a_H,t_H$
  can be expressed in a compact form as functions of two parameters 
$r_0/\mathcal{Q}$ and $r_0/r_h$,  where 
\begin{equation}
\mathcal{Q}=\sqrt{Q^2+Q_m^2} \ .
\end{equation}
Differently from the RN case,
the corresponding expression are involved and unenlightening and we shall
not include them here. They allow, nonetheless, for a fully analytical study of all solutions' properties. The charge to mass ratio is now defined as $q\equiv \mathcal{Q}/M$.

\begin{figure}[h!]
\begin{center}
\includegraphics[width=0.55\textwidth]{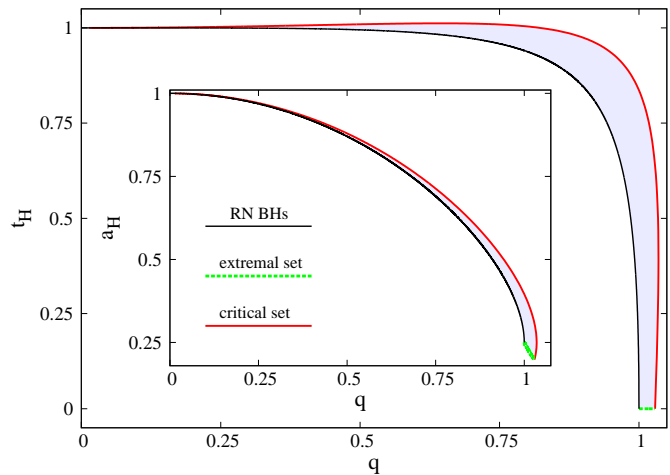}  
\caption{ 
The domain of existence of  RN-$F^4$ BHs
with $Q=Q_m$ in a reduced temperature (main panel) or reduced area (inset) $vs.$ charge to mass ratio, $q$, plot. 
}
\label{domainF4}
\end{center}
\end{figure}

The domain of existence of these solutions with equal electric and magnetic charges 
is shown in Fig.~\ref{domainF4}.
One observes the existence of overcharged solutions, $i.e.$ with $q>1$ , for some range of parameters.
Another remark is that the minimal value of the reduced area $a_H$ is now $a_{min} = 0.2$,
a value attained for $q=q_{max}=\simeq 1.0278$.
The domain of existence   is 
 bounded by three curves:
i) the RN limit ($i.e.$ $\alpha \to 0$); ii) the extremal limit $T_H \to 0$
and iii) a critical set. 
This last set is a curious property which is specific to the considered model, 
and can be traced to the presence in~\eqref{f4psi} of $\arctan f(r)$ where
\begin{equation}
f(r)=
 \frac{\frac{r}{r_0}}{1-\frac{r^2}{2r_0^2} }  \ .
\end{equation}
This function possesses a pole at 
$r_c=\sqrt{2}r_0 $
which implies a different value of the function $\arctan f(r)$ 
as $r_c$ is approached from below or from above.
This results in a discontinuity of the metric functions $g_{rr}$ and $g_{tt}$
(as well as of the electric potential) at $r=r_c$.
Thus, to avoid this pathology to be manifest outside the horizon, we impose
\begin{eqnarray} 
r_h>r_c=\sqrt{2}r_0 \ ,
\end{eqnarray}
which results in the set of critical solutions shown in Fig. \ref{domainF4}.

Finally we remark that the Ricci scalar of the RN-$F^4$ solution is
\begin{eqnarray}
\label{R-our}
R=\frac{16 Q^2 r_0^4}{(r^4+4 r_0^4)^2}>0 \ ,
\end{eqnarray}
which implies that scalarisation can occur for 
$\xi<0$
only.

\subsection{The scalarised solutions}
\begin{figure}[h!]
\begin{center}
\includegraphics[width=0.55\textwidth]{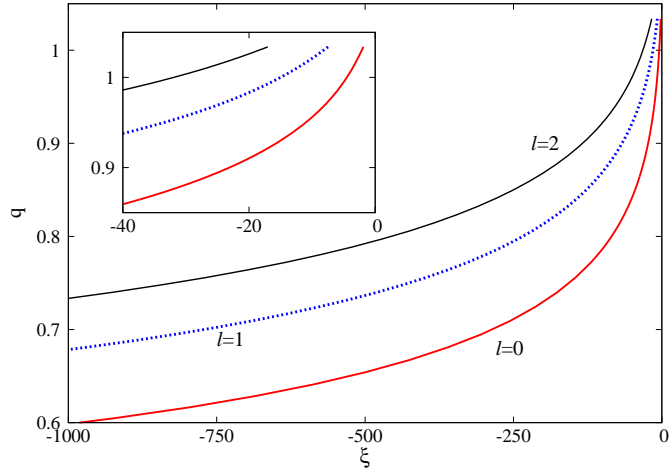}  
\caption{ 
$\ell=0,1,2$ existence lines for the RN-$F^4$
solution with $Q=Q_m=1$. The inset zooms in the region of small $|\xi|$.
}
\label{existence-lines-F4}
\end{center}
\end{figure}

We start, once more, the analysis by considering the zero modes, $i.e.$ the test field scalar clouds that occur at the onset of the scalarisation instability on the background of the dyonic BH described in the previous subsection. 
No exact solution appears to exist even in this limit  and the radial equation (\ref{zero-general}) is solved numerically.
 The corresponding results - the \textit{existence lines}  -  are shown in Fig.~\ref{existence-lines-F4}, for three different values of $\ell$. 
Amongst the notable features we highlight: 
$(i)$ the existence of a minimal value of $|\xi|$
below which no zero modes with a given number of nodes exist; 
$(ii)$ this minimal value of $|\xi|$
increases with $\ell$; 
and
$(iii)$ when varying $\xi$, the  zero modes exist for all range of $q=\mathcal{Q}/M$, 
with $\xi\to -\infty$ as $q\to 0$.

\begin{figure}[h!]
\begin{center}
\includegraphics[width=0.55\textwidth]{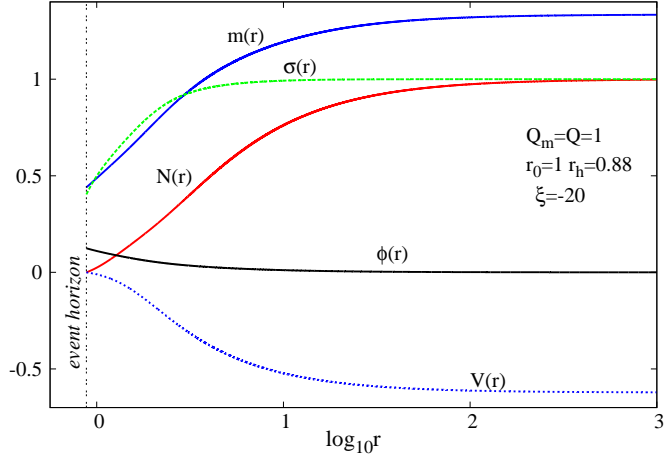}  
\caption{ 
The profile of a typical scalarized RN-$F^4$ solution.
}
\label{profileRNF4}
\end{center}
\end{figure}

 The construction of the nonlinear continuation of the scalar clouds follows directly.
 The far field asymptotics of the 
 scalarised solutions are similar,  at leading order,  to those the scalar-free solution (\ref{newF4}):  $N(r)\to 1$,  $\sigma(r)\to 1$,
$V(r)\to \Phi+Q/r$
and $\phi(r)\to 0$, as $r\to \infty$.
The solution will also posses a horizon at $r=r_h>0$,
where
$N(r_h)=0$, $V(r)=0$ and   $\sigma(r)$ is strictly positive.
Then we are left with a system of  four non-linear ordinary differential equations (plus a constraint)
for the functions $N,\sigma$, $\phi$ and $V(r)$.
We note that the equation for $V(r)$ possesses the first integral
\begin{eqnarray}
\label{fint}
V'=-\frac{Q r^2 \sigma}{r^4+4r_0^4}\ .
\end{eqnarray} 
which allows to treat $Q$
as an input parameter.
Then the  scalarised BHs are found by solving
numerically  two first order equations for $N,\sigma$ and a second order equation for $\phi$.
The profile of a typical solution is shown in Fig.~\ref{profileRNF4}.

\begin{figure}[ht!]
\begin{center}
{\includegraphics[width=8cm]{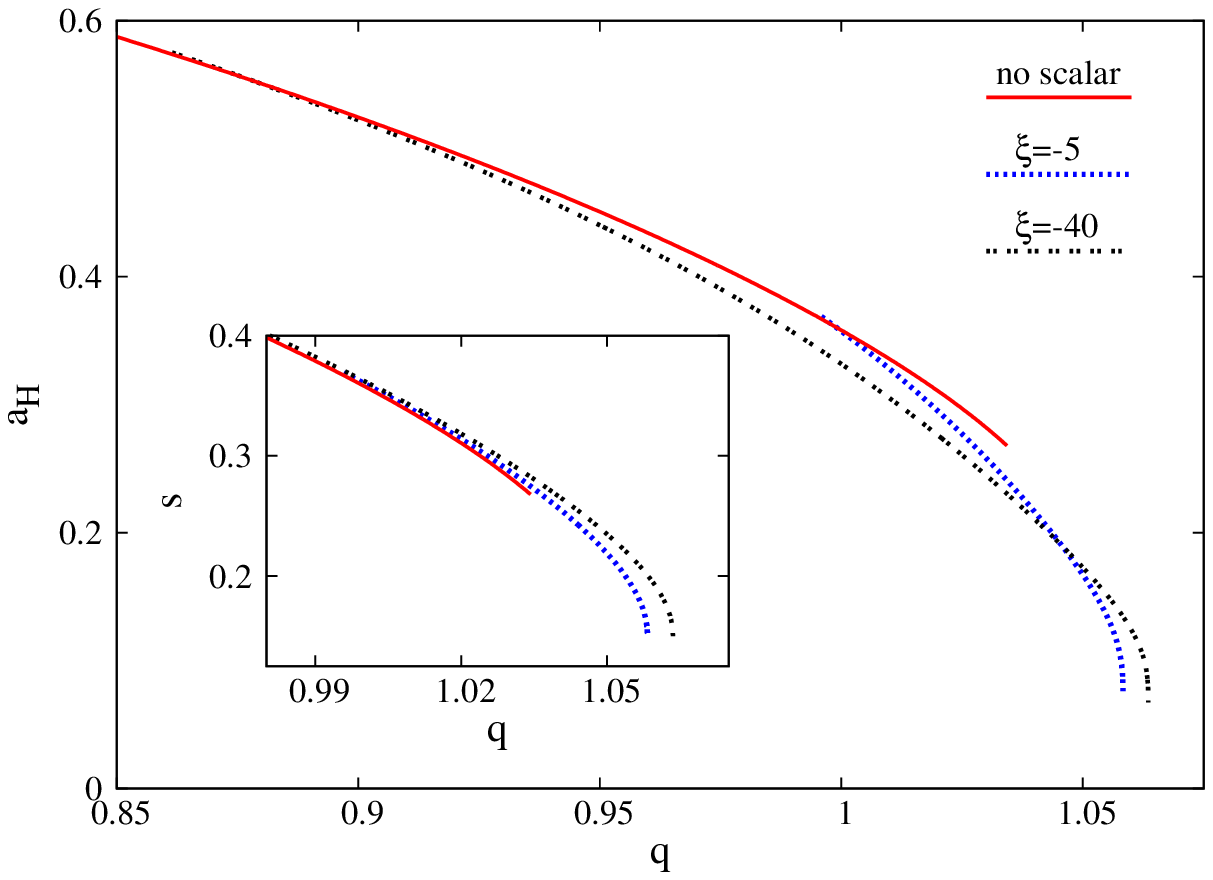}}
{\includegraphics[width=7.8cm]{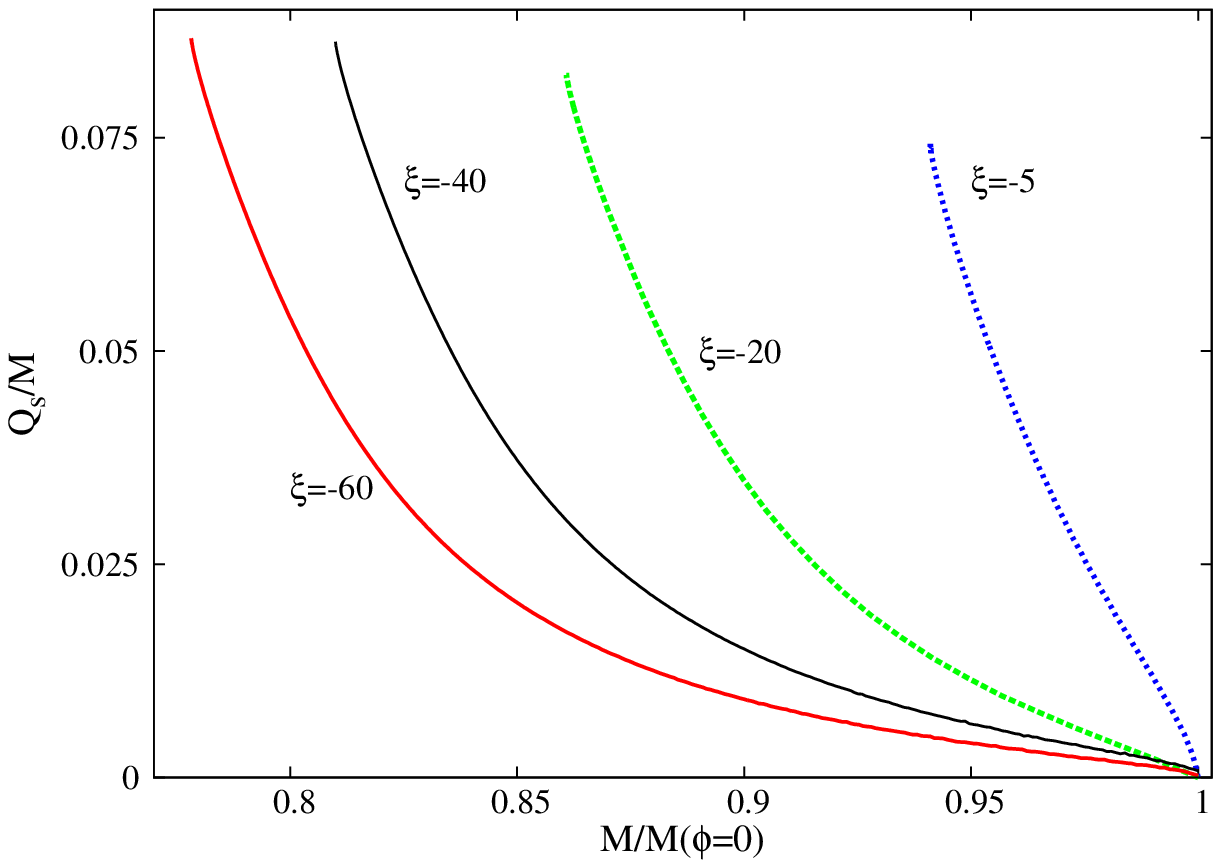}}
\caption{
(Left panel) Reduced area (main panel) and reduced entropy (inset) in terms of the charge to mass ratio $q$ for RN-$F^4$ BHs and their scalarised counterparts for two values of the coupling $\xi$. Observe that for fixed $q$ the scalarised solutions are entropically favoured. 
(Right panel) Reduced scalar `charge' in terms of the ADM mass normalised to the mass at the bifurcation point, for four values of the coupling. 
}
\label{sF4}
\end{center}
\end{figure}

The scalarised solutions still have three
global charges $(M,Q,Q_m)$
and, in the region of co-existence,
 have a larger entropy than the corresponding solutions with $\phi=0$, even though they can have a smaller horizon area - $cf.$ Fig.~\ref{sF4}. This figure also shows that one finds, as in the GHS case, overcharged solutions with $q>1$. In the right panel of Fig.~\ref{sF4} one observes that, for larger coupling $|\xi|$ the reduced scalar charge grows more slowly in terms of the mass of the solution normalised to that of the bifurcating point. Finally, we notice that the limiting behaviour of the branches of scalarised BHs with fixed $\xi$
seems to be similar to that found in the scalar-free case, with the existence of an  
 extremal limit which is regular.

\section{NCG inspired modified Schwarzschild BHs}
\label{sec6}

\subsection{The scalar-free solutions}

As our final example, we consider the NCG inspired Schwarzschild BH,  the NSS solution~\cite{Nicolini:2005vd}. One may regard this example as taking into account the effects of quantum gravity, in the non-commutative geometry approach, yielding a solution which is not scale invariant. The solution solves the Einstein equations (\ref{Einstein-eqs})
with an energy-momentum tensor having the following non-vanishing components
\begin{eqnarray}
\label{source}
T_r^r=p_r(r)\ , \qquad T_\theta^\theta =p_\theta(r)\ , \qquad T_t^t=-\rho(r) \ ,
\end{eqnarray}
where 
\begin{eqnarray}
 p_r(r) =-\rho(r) \ .
\end{eqnarray}
Assuming spherical symmetry and taking the metric ansatz
(\ref{gen-metric}) with the choices (\ref{gc}), 
the conservation of this energy-momentum tensor, $T_{\mu;\nu}^\nu=0$ implies
\begin{eqnarray}
 p_\theta(r) =-\rho(r)-\frac{r \rho'(r)}{2} \ .
\end{eqnarray}

The NSS solution is obtained postulating a smeared, particle-like source with
\begin{eqnarray}
 \rho(r)=\frac{M}{2\sqrt{\pi }\vartheta^{3/2}}e^{-\frac{r^2}{4\vartheta}} \ ,
\end{eqnarray}
where $\vartheta$ is  an input parameter of the theory related to the spacetime non-commutativity, which introduces a new length scale $r_0$:\footnote{This choice simplifies a number of relations below. We work in units with $4\pi G=1$ (while the choice in \cite{Nicolini:2005vd} was $G=1$).}
\begin{eqnarray}
 \vartheta =\frac{r_0^2}{4} \ .
\end{eqnarray}
Then the Einstein equations (\ref{Einstein-eqs}) yield the following expressions the metric functions in
(\ref{gen-metric})
\begin{eqnarray}
\label{NSS1}
\sigma(r)=1\ , \qquad N(r)=1-\frac{4M}{r\sqrt{\pi}}\gamma\left(\frac{3}{2},\frac{r^2}{r_0^2}\right) \ ,
\end{eqnarray}
where $\gamma$ is the lower incomplete Gamma function
\begin{eqnarray}
\gamma\left(\frac{3}{2},\frac{r^2}{r_0^2}\right)=\int_0^{r^2/r_0^2}dt \,  t^{1/2}e^{-t} \ .
\end{eqnarray}
This solutions possess an (outer) horizon at the  $N(r_h)=0$, which fixes the relation between the ADM mass and horizon radius,
\begin{eqnarray}
M=\frac{\sqrt{\pi} r_h}{4\gamma(\frac{3}{2},\frac{r_h^2}{r_0^2})}~.
\end{eqnarray}
The Hawking temperature and the horizon area of the solutions are
\begin{eqnarray}
T_H=\frac{1}{4\pi r_h}
\left\{
1+\frac{2r_h^2}{r_0^2}\ \left[1-\frac{r_0}{2r_h}e^{r_h^2/r_0^2}\sqrt{\pi}{\rm erf}\left(\frac{r_h}{r_0}\right)\right]^{-1}
\right\} \ , \qquad A_H=4\pi r_h^2 \ ,
\end{eqnarray}
where ${\rm erf}$ is the error function. We remark that BHs become cold both in the limit of very large mass (classical limit) and in a new limit where they become extremal $(T_H=0)$, which occurs for $r_h=r_c\simeq 1.51122 r_0$ ($i.e.$ with $A_H\neq 0$) - see Fig. 4~\cite{Nicolini:2005vd}.

Finally,  the expression of the Ricci scalar is
\begin{eqnarray}
R=-\frac{16 e^{-\frac{r^2}{r_0^2}}M(r^2-2r_0^2)}{\sqrt{\pi}r_0^5}<0 \ ,
\end{eqnarray}
which implies
that scalarisation occurs  in this case for $\xi>0$.

\subsection{The scalarised solutions}
\begin{figure}[h!]
\begin{center}
\includegraphics[width=0.55\textwidth]{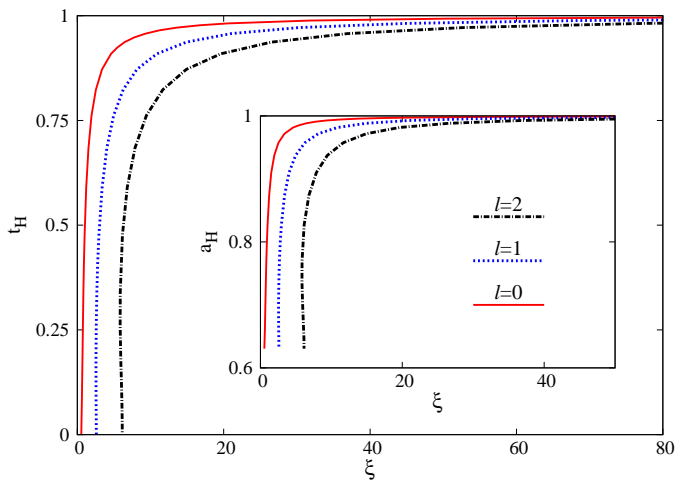}  
\caption{ 
$\ell=0,1,2$ existence lines for the NSS BHs.
}
\label{zero-mode-N}
\end{center}
\end{figure}

The study of these scalarised solutions is done following closely the approach in the previous two examples.
Starting again with the case of an infinitesimally small scalar field, the radial equation~(\ref{zero-general}) is solved  
for a NSS background.
The corresponding existence lines are shown in Fig.~\ref{zero-mode-N} 
for $\ell=0,1,2$, in a reduced temperature (or area) $vs.$ coupling diagram.
 Again, scalar clouds exist for any  NSS BH, provided
one considers a particular set of coupling constants $\xi_n$, where $n$ is the node number, although here we always take $n=0$ and denote $\xi_0\rightarrow \xi$.
One observes that, for a given ADM mass, 
the value of $\xi$ increases with the Hawking temperature. Approaching the classical (Schwarzschild) limit $(r_0\rightarrow 0$) scalarisation
requires $\xi \to \infty$.

The non-linear  continuation of the scalar zero modes
has a subtlety
related to the fluid source 
(\ref{source}). This source contains a constant fixing the mass of the solutions.  
In our approach, however, we solve the Einstein equations with
an energy-momentum tensor which is the sum of the fluid tensor (\ref{source})
plus the scalar field contribution (\ref{Tiks}).
For a given model with a specific $\xi$,
the constant $M$ in  the fluid expression
(\ref{source}) is fixed to be $M=M(\phi=0)$,
$i.e.$ the value of the scalar-free solution at the bifurcation point. 

\begin{figure}[h!]
\begin{center}
\includegraphics[width=0.55\textwidth]{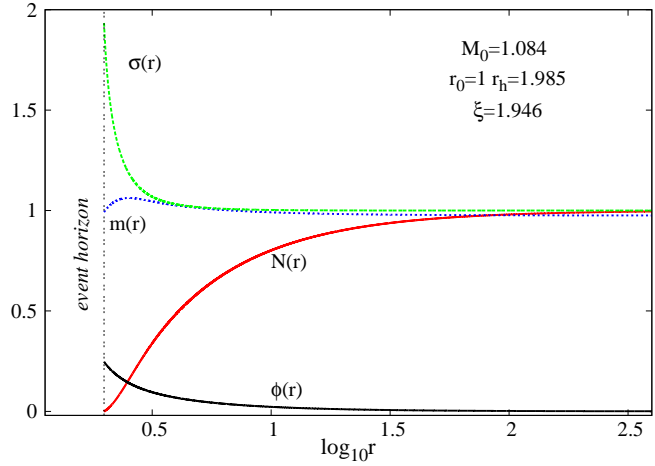}  
\caption{ 
The profile functions of a typical scalarised NSS solution.
}
\label{profileNSS}
\end{center}
\end{figure}

A typical scalarised solution is shown in Fig.~\ref{profileNSS}.
One can see that, different from the other two cases discussed above,
$m'<0$ for large enough values of $r$, which corresponds to occurrence of a region
with negative energy density which extends to infinity, although the total mass is still positive for all solutions.

\begin{figure}[ht!]
\begin{center}
{\includegraphics[width=8cm]{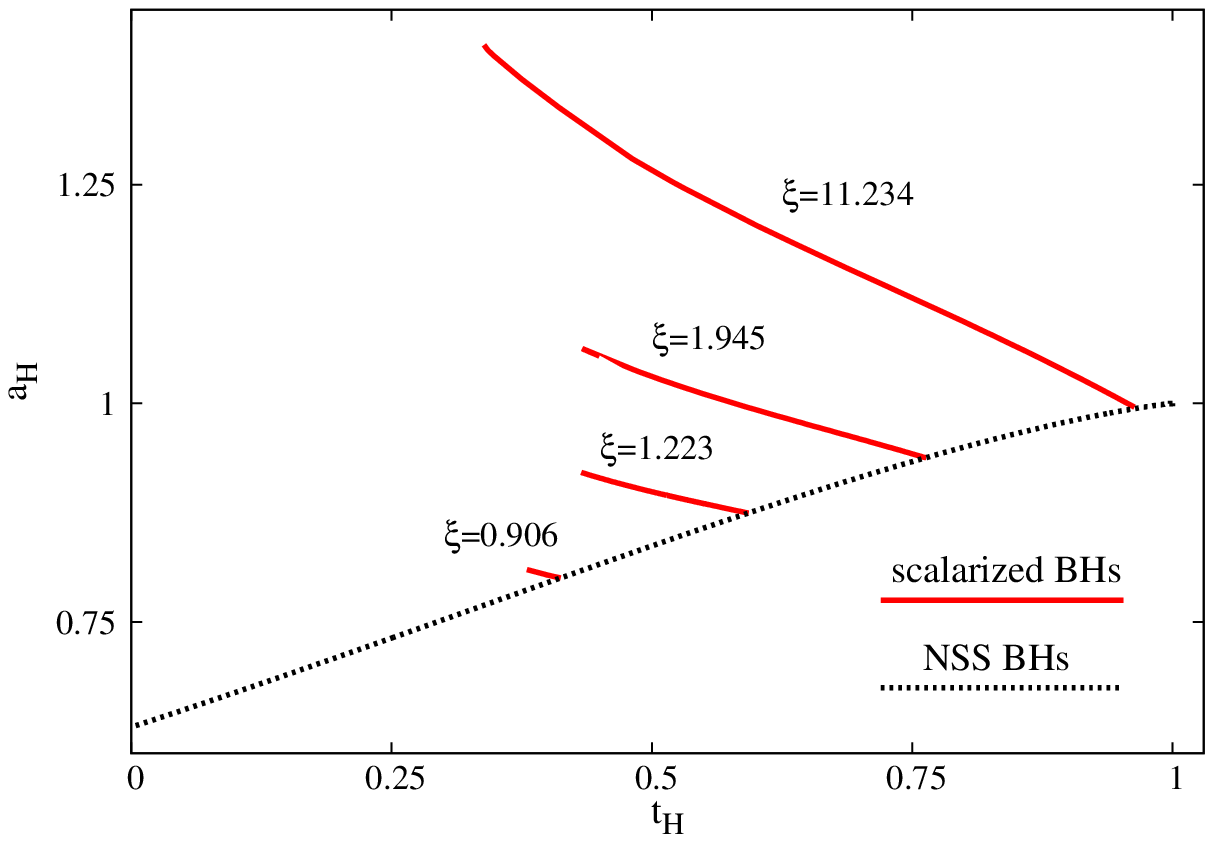}}
{\includegraphics[width=7.8cm]{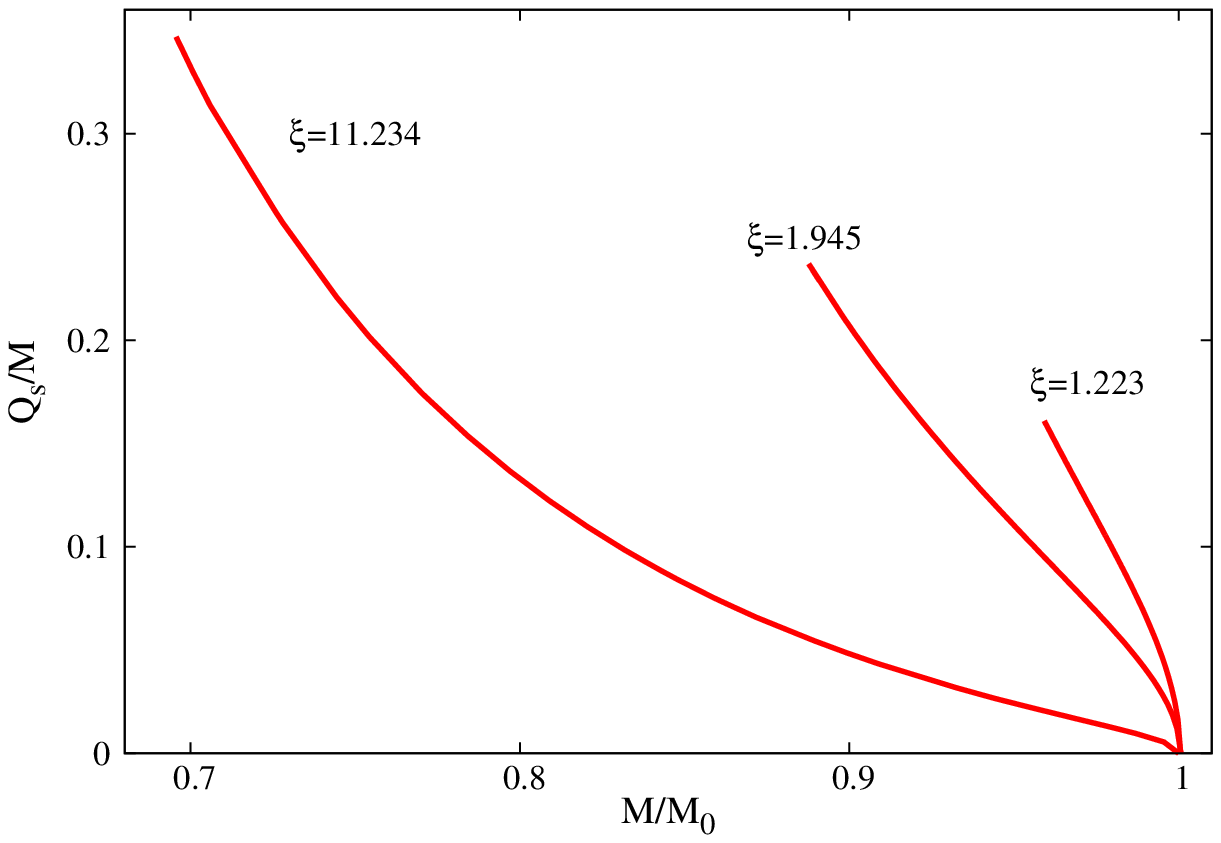}}
\caption{
(Left panel) Branching off of the scalarised solutions from the NSS trunk in a reduced area $vs.$ reduced temperature diagram for different values of $\xi$. (Right panel) Reduced scalar charge as a function of the ADM mass of the scalarised NSS BHs normalised to the mass of the bifurcation point. 
}
\label{NSScsi}
\end{center}
\end{figure} 

In Fig.~\ref{NSScsi} (left panel) the bifurcation of the scalarised solutions from the trunk of NSS solutions is shown in a reduced area $vs.$ reduced temperature diagram, for several values of the coupling $\xi$. One observes that fixing the temperature the scalarised BH has always a larger reduced area. For lower temperatures, as the NSS solution is approaching extremality scalarisation requires a smaller coupling. One can observe that the extent of the scalarised branches 
decreases with  $\xi$. In the right panel of Fig.~\ref{NSScsi}  one observes that, again, for smaller couplings the reduced scalar charge increases faster with the mass of the solutions normalised to that of the bifurcation point.
Finally, one remarks that, differently from the other two case above, the answer to the question of which solution maximises the entropy for given global charges  is not unambiguous. 
Firstly, the consensus in the literature~\cite{Nicolini:2008aj} 
is that the quantum gravity corrections give a correction to the Bekenstein-Hawking formula
 already in the scalar-free NSS case (which is computed by integrating the 1st law).
A further complication is due to the existence of a `background' fluid source, which provides an extra-parameter for the scalarised solutions.
Nonetheless, the existence of a spontaneous scalarisation instability of the NSS background, 
which is clear at the linear level, suggests that there should be entropically preferred scalarised solutions.

\section{Further remarks}
\label{sec7}
In this paper we have discussed the possibility that BH solutions which are not scale invariant, in the sense their energy-momentum tensor is traceful, can scalarise in the presence of a non-minimal coupling between a the scalar field and the Ricci scalar curvature. Here we have considered this scalarisation in the context of quantum field theory motivated non-minimal coupling $\xi \phi^2 R$, but similar conclusions can be extracted for more general scalar-tensor theories. 

We have considered three illustrative examples of non-scale-invariant BHs. In two of the considered cases the breakdown of scale invariance can be attributed to quantum effects (a sort of trace anomaly) whereas in the remaining case it is due to the presence of scale-invariance braking classical matter. In all cases the pattern is similar. The scalar-free solutions become prone to a tachyonic instability in some region of the parameter space. In particular at the onset of the instability there is a zero mode (scalar cloud) that we have computed in all examples. Then scalarised solutions branch off from the scalar-free ones at bifurcating points corresponding to the latter solutions that can suppor the scalar clouds. Since the scalarised solutions are entropically preferred (in the cases the entropy is unambiguous) for the same global charges it seems reasonable to anticipate that they will be the endpoint of the instability observed for the scalar free solutions. This was dynamically confirmed in~\cite{Herdeiro:2018wub} for another model of matter-induced spontaneous scalarisation. Establishing a similar result in the present cases also requires fully non-linear numerical simulations. 

Let us conclude with some possible further avenues of related research. Firstly, other features of the scalarised BHs unveiled in~\cite{Herdeiro:2018wub} should occur also in this case. For example, we predict the existence of static BHs {\it without isometries}
also for this type of scalarisation, 
branching off from the scalar-free trunck at zero-modes with higher $\ell$ and azimuthal quantum number $m\neq 0$. 
Secondly, it would be interesting to further investigate the scalarised BHs for 
other, potentially more realistic generalisations of Schwarzschild BHs, within quantum gravity frameworks. Thirdly, one could consider higher dimensions and different asymptotics. 
In particular, concerning (A)dS asymptotics, BHs with scalar hair are known to exist, being supported by the $\xi \phi^2 R$ term~\cite{Winstanley:2002jt,Radu:2005bp}.
It would be interesting to reconsider them in the context of spontaneous scalarisation. 
Finally, It would be interesting
to investigate the connection of the scalarisation mechanism
discussed in this work with the quantum instabilities due to a non-minimally coupling  (`awaking the vacuum')
discussed, $e.g.$ in~\cite{Lima:2010na,Landulfo:2012nz,Mendes:2013ija}.

\bigskip

{\bf Acknowledgements}
\\
E.R. thanks D.H. Tchrakian for valuable discussions and for collaboration
on related projects.
 This work has been supported by the FCT (Portugal) IF programme, by the FCT grant PTDC/FIS-OUT/28407/2017, by  CIDMA (FCT) strategic project UID/MAT/04106/2013, by CENTRA (FCT) strategic project UID/FIS/00099/2013 and
by  the  European  Union's  Horizon  2020  research  and  innovation  (RISE) programmes H2020-MSCA-RISE-2015
Grant No.~StronGrHEP-690904 and H2020-MSCA-RISE-2017 Grant No.~FunFiCO-777740. 
The authors would like to acknowledge
networking support by the
COST Action CA16104. 
E.R. is grateful for hospitality and support to Dublin Institute for Advanced Studies, where  a part of this work was done.


\end{document}